\documentclass[letterpaper,english,prl,aps,twocolumn,10pt,longbibliography,superscriptaddress]{revtex4-1}

\usepackage{amsmath,amssymb,amsfonts}
\usepackage{graphicx}
\usepackage{enumerate}

\pdfpageheight\paperheight
\pdfpagewidth\paperwidth

\usepackage{color}
\definecolor{BLACK}{gray}{0}
\definecolor{WHITE}{gray}{1}
\definecolor{RED}{rgb}{1,0,0}
\definecolor{GREEN}{rgb}{0,1,0}
\definecolor{BLUE}{rgb}{0,0,1}
\definecolor{CYAN}{cmyk}{1,0,0,0}
\definecolor{MAGENTA}{cmyk}{0,1,0,0}
\definecolor{YELLOW}{cmyk}{0,0,1,0}
 
\usepackage[
colorlinks=true,linkcolor=red,citecolor=green,urlcolor=blue,
pdfstartview=FitH,
bookmarksopen=true,bookmarksopenlevel=0,
plainpages=false,pdfpagelabels,
pagebackref=false,
pdftoolbar=false]{hyperref}

\begin{document}

\title{Designing a practical high-fidelity long-time quantum memory}

\author{Kaveh Khodjasteh}
\affiliation{\mbox{Department of Physics and Astronomy, Dartmouth 
College, 6127 Wilder Laboratory, Hanover, NH 03755, USA}}

\author{Jarrah Sastrawan}
\author{David Hayes}
\author{Todd J. Green}
\author{Michael J. Biercuk}

\affiliation{\mbox{ARC Centre for Engineered Quantum Systems, School
of Physics, The University of Sydney, New South Wales 2006,
Australia}}

\author{Lorenza Viola}
\thanks{Corresponding author (lorenza.viola@dartmouth.edu)}
\affiliation{\mbox{Department of Physics and Astronomy, Dartmouth 
College, 6127 Wilder Laboratory, Hanover, NH 03755, USA}}

\begin{abstract}
Quantum memory is a central component for quantum information
processing devices, and will be required to provide high-fidelity storage of
arbitrary states, long storage times, and small access latencies.  Despite 
growing interest in applying physical-layer error-suppression strategies to 
boost fidelities, it has not previously been possible to meet such competing demands with a single approach.  
Here, we use an experimentally validated theoretical framework to identify periodic repetition of a
high-order dynamical decoupling sequence as a systematic strategy to
meet these challenges.  We provide analytic bounds -- validated by numerical calculations -- on the characteristics of the relevant control sequences and show that a ``stroboscopic saturation" of coherence, or \emph{coherence plateau}, 
can be engineered, even in the presence of experimental imperfection.  This permits high-fidelity storage for times that can be exceptionally long, meaning that our device-independent results should prove instrumental in producing
practically useful quantum technologies.
\end{abstract}

\date{\today}
\maketitle

Developing techniques for the preservation of arbitrary quantum states
-- that is, quantum memory -- in realistic, noisy physical systems is
vital if we are to bring quantum-enabled applications including secure
communications and quantum computation to reality.  While numerous
techniques relying on both open- and closed-loop control have
been devised to address this challenge, {\em dynamical error
  suppression} strategies based on dynamical decoupling (DD)
\cite{Viola1998,Viola1999Dec,khodjasteh_fault-tolerant_2005,Uhrig2007}, dynamically
corrected gates (DCGs) \cite{khodjasteh_dynamically_2009,khodjasteh_arbitrarily_2010}, and composite
pulsing~\cite{True} are emerging as a method of choice for {\em
physical-layer} decoherence control in realistic settings described
by non-Markovian open-quantum-system dynamics.  Theoretical and
experimental studies in a variety of platforms
\cite{BiercukNature,UysPRL,NirPRL,NirJMP,Foletti,Bluhm,Schulman,Barthel,Medford,Lyon,Wang,deLange,Cory,Naydenov,WangNV,hayes_dcg} have consistently
pointed to dynamical error suppression as a resource-efficient
approach to substantially reducing physical error rates.

Despite these impressive advances, investigations to date have largely
failed to capture the typical operating conditions of any true quantum
memory; namely, high-fidelity storage of quantum information for {\em
arbitrarily long storage times}, with on-demand access.  This would be
required, for instance, in a quantum repeater, or in a quantum
computer where some quantum information must be maintained with error
rates deep below fault-tolerant thresholds while large blocks of an
algorithm are carried out on other qubits.  Instead, both experiment
and theory have primarily focused on two control
regimes~\cite{MikeFF}: the ``coherence-time regime,'' where the goal
is to extend the characteristic (``$1/e$'' or $T_2$) decay time for
coherence as long as possible, and the ``high-fidelity regime,'' where
the goal is to suppress errors as low as possible for storage times
short compared to $T_2$ (for instance, during a single gating period).
Similarly, practical constraints on control timing and access latency
-- of key importance to laboratory applications -- have yet to be
considered in a systematic way.

In this Article, we demonstrate how to realize a practically useful
quantum memory via dynamical error suppression.  Specifically, our
studies identify the \emph{periodic repetition of a high-order DD
sequence} as an effective strategy for memory applications,
considering realistic noise models, incorporating essential
experimental limitations on available controls, and addressing the key
architectural constraint of maintaining short access latencies to
stored quantum information.  We consider a scenario where independent
qubits couple to a noisy environment, 
and both dephasing and depolarization errors introduced by realistic
DD sequences of bounded-strength $\pi$-pulses are fully accounted for.
We analytically and numerically characterize the achievable long-time
coherence for repeated sequences and identify
conditions under which a stroboscopic ``coherence plateau'' can be
engineered, and fidelity \emph{guaranteed} to a desired level at long
storage times -- even in the presence of experimentally realistic
constraints and imperfections. We expect that our approach will
provide a practical avenue to high-fidelity low-latency quantum
storage in realistic devices.

\vspace*{1mm}

\noindent 
{\bf Results}\\
\noindent 
{\bf Model.}  The salient features of our approach may be appreciated
by first focusing on a single qubit subject to dephasing.  In the absence of control, we consider 
a model Hamiltonian of the form $H\equiv \sigma_z \otimes (\epsilon_0
+ B_z) + H_E$, where the Pauli matrix $\sigma_z$ and $\epsilon_0$
define the qubit quantization axis and internal energy, respectively
(we can set $\epsilon_0=0$ henceforth), and $B_z$, $H_E$ are operators
acting on the environment Hilbert space. An {\em exact} analysis of
both the free and the controlled dynamics is possible if the
environment can be described in terms of either a quantum bosonic bath
in thermal equilibrium (spin-boson model), a weakly-coupled quantum
spin bath (spin-bath model), or a stationary Gaussian stochastic
process (classical-noise model)
\cite{Palma,Viola1998,Uhrig2007,Cyw2008,Hodgson2010,khodjasteh_limits_2011,HKBV2011,Sasaki,Suter2}.
Such dephasing models provide an accurate physical description
whenever relaxation processes associated with energy exchange occur
over a characteristic time scale ($T_1$) substantially longer than any
typical time scale associated with the dephasing dynamics.  As a result,
our analysis is directly relevant to a wide range of experimentally
relevant qubit systems, from trapped ions and atomic ensembles
\cite{BiercukNature,NirPRL} to spin qubits in nuclear and electron magnetic
resonance and quantum dots~\cite{Suter1,Suter2,Lyon,Foletti,Bluhm,Schulman}.

We shall proceed by considering the effects of DD within a
filter-design framework which generalizes the transfer-function
approach widely used across the engineering community \cite{Rutman}
and provides a transparent and experimentally relevant picture of the
controlled dynamics in the frequency domain~\cite{Kurizki2001,
Cyw2008,BiercukNature,UysPRL,MikeFF,GreenFF}.  In order to more easily introduce
key concepts and clearly reveal our underlying strategy, we first
consider an idealized ``bang-bang'' DD setting in which perfect
instantaneous $\pi$ rotations are effected by using unbounded control
amplitudes.  As we move forward, we will relax these unphysical
constraints, and demonstrate how similar results may be obtained with
experimentally realistic controls.

\begin{figure}[b]
\includegraphics[width=0.90\columnwidth]{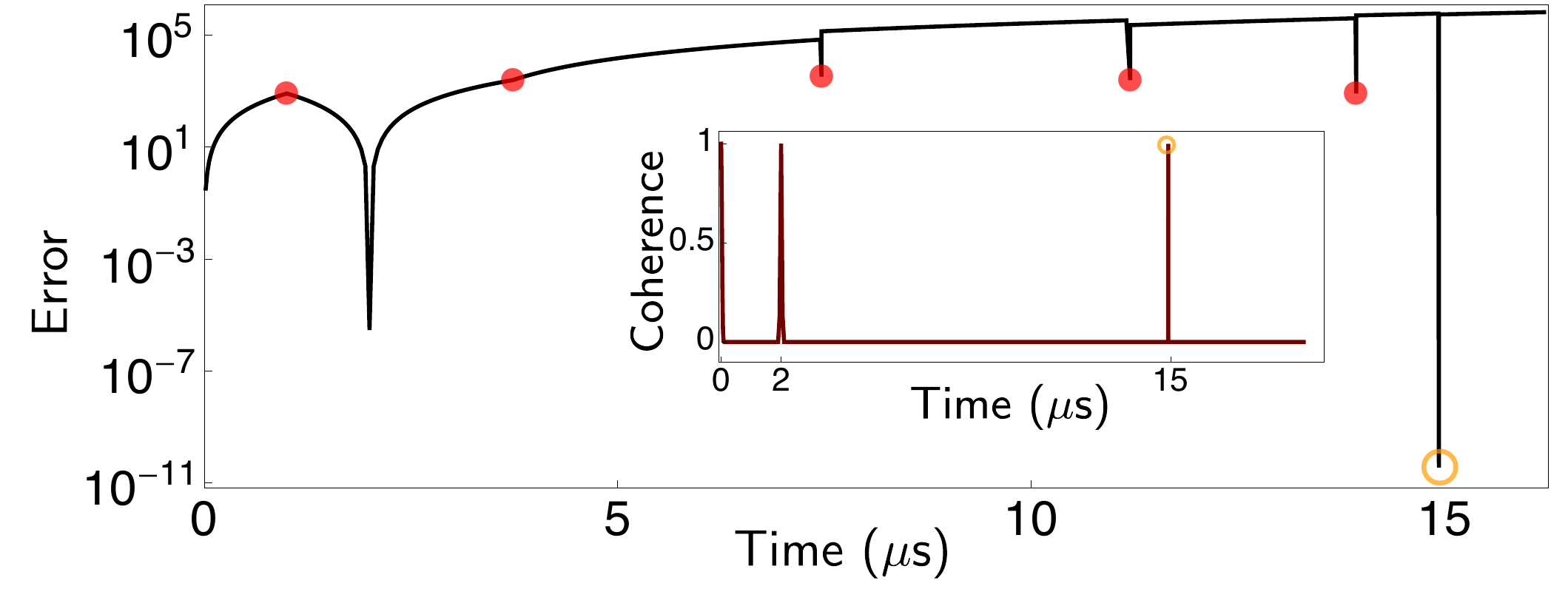} 
\vspace*{-2mm}
\caption{{\bf Access-latency in high-order DD sequences.}  DD error
and coherence (inset) \emph{during} a UDD$_5$ sequence with minimum
interpulse time $\tau=1\,\mu$s.  Pulse times are marked with filled
circles while the open circle indicates the readout time $T_p$.
Minimal error (maximal coherence) is reached only at the conclusion of
the sequence, with the coherence spike near 2$\,\mu$s resulting from a
spin-echo effect.  For illustration purpose, in all figures we assume a
phenomenological noise model appropriate for nuclear-spin induced
decoherence in a spin qubit in GaAs, $S(\omega)= g
(\omega/\omega_c)^{-2}e^{-\omega^2/\omega_{c}^{2}}$, with $\omega \in
[\omega_{\text{min}},\omega_{\text{max}}]$. We set $g =0.207
\omega_{c}$, $\omega_{c}/2\pi =10\text{kHz}$,
$\omega_{\text{min}}/2\pi=0.01\:$Hz, and
$\omega_{\text{max}}/2\pi=10^8\:$Hz to maximize agreement with the
measured $T_2$ ($\approx$ 35 ns) \cite{Bluhm,BiercukBluhm}.  We chose
$\tau$ {\em well above} technological constraints ($\sim$ ns) in order
to reduce $n$.
\label{fig:latency}}
\end{figure}

In such an idealized control scenario, a DD sequence may be specified
in terms of the pulse-timing pattern $p\equiv \{t_j \}_{j=1}^n$,
where we also define $t_{0}\equiv 0$, $t_{n+1}\equiv T_{p}$ as the
sequence duration, and we take all the interpulse intervals ($t_{j+1}
- t_j$) to be lower-bounded by a \emph{minimum interval}
$\tau$~\cite{khodjasteh_limits_2011}.  The control propagator reads
$U_{c}(t)= \sigma_x^{[y_p(t)+1]/2}$, with $y_{p}(t)$ being a
piecewise-constant function that switches between $\pm 1$ whenever a
pulse is applied. The effect of DD on qubit dephasing may be evaluated
exactly in terms of a spectral overlap of the control modulation and
the noise power spectral density,
$S(\omega)$~\cite{Kurizki2001,Cyw2008}, which is determined by the
Fourier transform of the two-time noise correlation function
\cite{Sasaki}.  Typically, $S(\omega)$ has a power-law behavior at low
frequencies, and decays to zero beyond an upper cutoff $\omega_c$,
that is, $S(\omega) \propto \omega^s f(\omega , \omega_c)$, and the
``rolloff function'' $f$ specifies the high-frequency behavior,
$f=\Theta(\omega-\omega_c)$ corresponding to a ``hard'' cutoff.  Let
$\tilde{y}_{p}(\omega)$ denote the Fourier transform of $y_{p}(t)$,
which is given by $\tilde{y}_{p}(\omega)=\omega^{-1}
\sum_{j=0}^{n}(-1)^{j}[\exp(it_{j}\omega)-\exp(it_{j+1}\omega)]$
\cite{Uhrig2007,Cyw2008}. The \emph{filter function} (FF) of the
sequence $p$ is given by
$F_{p}(\omega)=\omega^{2}\left|\tilde{y}_{p}(\omega)\right|^{2}$, and
the bang-bang-controlled qubit coherence decays as $e^{-\chi_{p}}$,
where the \emph{decoupling error} $\chi_{p}= \int_{0}^{\infty}
\frac{S(\omega)}{2 \pi \omega^2} \, F_{p}(\omega)\, d\omega$ at time
$t=T_p$, and the case $n=0$ recovers free evolution over $[0, T_p]$.

In this framework, the applied DD sequence behaves like a
``high-pass'' filter, suppressing errors arising from slowly
fluctuating (low-frequency) noise.  Appropriate construction of the
sequence then permits the bulk of the noise power spectrum to be
efficiently suppressed, and coherence preserved.  For a given sequence
$p$, this effect is captured quantitatively through the \emph{order of
error suppression} $\alpha_p$, determined by the scaling of the FF
near $\omega=0$, that is, $F_{p}(\omega) \equiv |A_{\text{bb}}|^2
\omega^{2(\alpha_{p}+1)}\propto (\omega\tau)^{2(\alpha_{p}+1)}$, for a
sequence-dependent pre-factor $A_{\text{bb}}$. A high multiplicity of
the zero at $\omega=0$ leads to a perturbatively small value of
$\chi_p$ as long as $\omega_{c}\tau\ll 1$.  In principle, one may thus
achieve low error probabilities over a desired storage time $T_s$
simply by using a high-order DD sequence, such as Concatenated DD
(CDD,~\cite{khodjasteh_fault-tolerant_2005}) or Uhrig DD
(UDD,~\cite{Uhrig2007}), with the desired storage time $T_s \equiv
T_p$.

\vspace*{2mm}

\noindent
{\bf Quantum memory requirements.}  Once we attempt to move beyond
this idealized scenario in order to meet the needs of a practically
useful, long-time quantum memory, several linked issues arise.
First, perturbative DD sequences are {\em not} generally viable for
high-fidelity long-time storage as they require arbitrarily fast
control ($\tau\rightarrow0$).
Real systems face systematic constraints mandating $\tau >0$, and as a
result, increasing $\alpha_{p}$ necessitates extension of $T_{p}$,
placing an upper bound on high-fidelity storage
times~\cite{Hodgson2010,UhrigLidar,khodjasteh_limits_2011}.  
(For instance, a UDD$_n$ sequence achieves $\alpha_p=n$ with $n$ pulses, applied at $t_j = T_p \sin^2 [\pi j/(2 n + 2)]$. For fixed $T_p$, increasing $\alpha_p$ implies increasing $n$, at the expenses of shrinking $\tau$ as $\tau \equiv t_1 ={\cal O}(T_p/n^2)$. If $\tau>0$ is fixed, and $\alpha_p$ is increased by lengthening $T_p$, eventually the perturbative corrections catch up, preventing further error reduction.)
Second, potentially useful numerical DD approaches, such as randomized DD \cite{Viola2005Random,Santos2006} or
optimized ``bandwidth-adapted'' DD \cite{khodjasteh_limits_2011},
become impractical as the configuration space of {\em all} possible DD
sequences over which to search grows exponentially with $T_s$.
Third, DD exploits interference pathways between control-modulated
trajectories, meaning that mid-sequence interruption ($t<T_{p}$)
typically result in significantly sub-optimal performance
(Fig. \ref{fig:latency}). However, a stored quantum state in a
practical quantum memory must be accessible not just at a designated
final retrieval time but at intermediate times also, at which it may serve as an input to a
quantum protocol.

Addressing all such issues requires a systematic approach to DD
sequence construction. Here, we identify a ``modular'' approach to
generate low-error, low-latency DD sequences for long-time storage out
of shorter blocks: {\em periodic repetition} of a base, high-order DD
cycle.

\vspace*{2mm}

\begin{figure}[tb]
\includegraphics[width=\columnwidth]{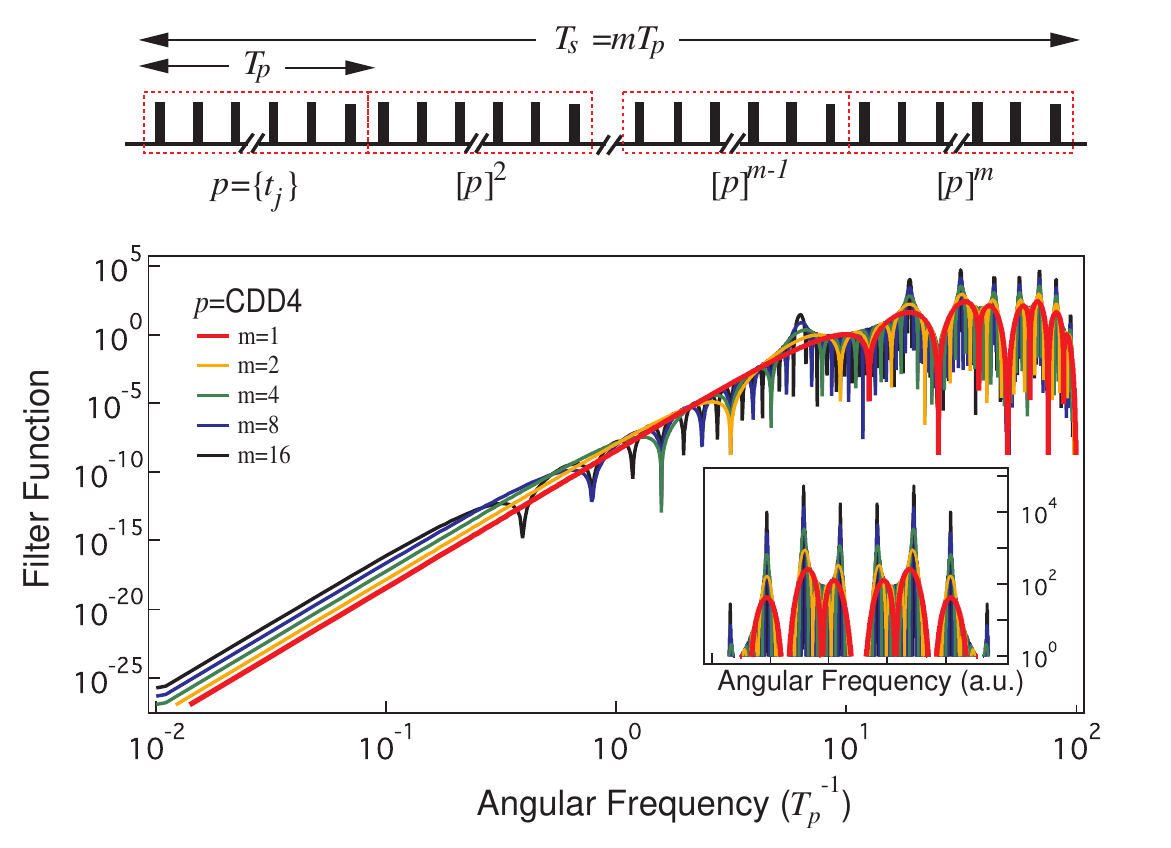} 
\vspace*{-5mm}
\caption{{\bf Schematic representation of base sequence repetition and
the effect on the filter function.}  Top: The base sequence $p$ is
indicated in red dashed boxes, and repeated $m$ times up to a total
storage time $T_{s}$.  Bottom: FF for repetition of a CDD$_4$ cycle.
The FF on a log-log plot grows with frequency with slope set by
$\alpha_{p}$ until it reaches the \emph{passband}, where noise is
passed largely unimpeded (red thick line).  Noise dominated by
spectral components in this region is efficiently suppressed by DD.
As $m$ grows, the sinusoidal terms in Eq.~(\ref{eq:contrep}) lead to
the emergence of ``resonance'' frequencies that modify the
single-cycle FF and produce sharp peaks in the passband.  These must
be considered when accounting for the effects of noise at long storage
time due to ``resonance'' effects.  Inset: FF passband on a log-linear
plot.
\label{fig:overview}}
\end{figure}

\noindent
{\bf Quantum memory via periodic repetition.}  The effect of
repetition for an \emph{arbitrary} sequence is revealed by considering
the transformation properties of the FF under sequence combination.
Consider two sequences, $p_1$ and $p_2$, joined to form a longer one,
denoted $p_1 + p_2$, with propagator $y_{p_1 + p_2}(t)$. In the
Fourier space we have $\tilde{y}_{p_{1}+ p_{2}}(\omega)=
\tilde{y}_{p_{1}}(\omega) + e^{i\omega
T_{p_1}}\tilde{y}_{p_{2}}(\omega).$ Let now $[p]^{m}$ denote the
sequence resulting from repeating $p$, of duration $T_{p}$, $m$ times,
with $T_s=m T_p$.  Computing $\tilde{y}_{[p]^m}(\omega)$ by iteration,
the following exact expression is found:
\begin{eqnarray}
\chi_{[p]^{m}} = \int_{0}^{\infty} \frac{S(\omega)}{2 \pi \omega^2}\,
\frac{\sin^2 (m\omega T_{p}/2)}{\sin^{2}(\omega T_{p}/2)}
F_{p}(\omega) \, d\omega .
\label{eq:contrep}
\end{eqnarray}
Equation~(\ref{eq:contrep}) describes dephasing dynamics under {\em
arbitrary} multipulse control, generalizing special cases in
which this strategy is implicitly used for simple base sequences
(periodic DD, $p=\{\tau ,\tau \}$~\cite{Hodgson2010} and Carr-Purcell,
$p=\{\tau,2\tau,\tau\}$), and showing similarities with the intensity
pattern due to an $m$-line diffraction grating \cite{Suter2}.  The
single-cycle FF, $F_p(\omega)$, is multiplied by a factor which is
rapidly oscillating for large $m$ and develops peaks scaling with
${\mathcal O}(m^2)$ at multiples of the ``resonance frequency,''
$\omega_\text{res} = 2\pi/T_p$, introduced by the periodic modulation
(see Fig. \ref{fig:overview} for an illustration).

After many repeats, the DD error is determined by the interplay
between the order of error suppression of the base sequence, the noise
power behavior at low frequencies, and the size of noise contributions
at the resonance frequencies.  The case of a hard upper frequency cutoff 
at $\omega_c$ is the simplest to analyze.  Applying the
Riemann-Lebesgue lemma removes the oscillating factor, resulting in
the following asymptotic expression:
\begin{equation} 
\lim_{m\rightarrow \infty} \chi_{[p]^m} \equiv
\chi_{[p]^{\infty}}=\int_{0}^{\omega_c}\frac{S(\omega)} {4 \pi
\omega^2}\,\frac{ F_{p}(\omega)} {\sin^{2}(\omega T_{p}/2)}d\omega,
\label{eq:plateau}
\end{equation}
provided that $\chi_{[p]^{\infty}}$ is finite.  The meaning of this
exact result is remarkable: for small $m$, the DD error initially
increases as $(m^2 \chi_p)$, until coherence stroboscopically
saturates to a non-zero {\em residual plateau} value
$(e^{-\chi_{[p]^{\infty}}})$, and \emph{no further decoherence
occurs}.  Mathematically, the emergence of this coherence plateau
requires that simple conditions be obeyed by the chosen base sequence
relative to the characteristics of the noise:
\begin{equation}
s+2\alpha_{p}>1,\;\; T_{p} \omega_{c}< 2 \pi,
\label{eq:crit}
\end{equation}
which correspond to removing the singularity of the integrand in
Eq. (\ref{eq:plateau}) at $0$ and $\omega_{\text{res}}$, respectively.
Thus, judicious selection of a base sequence, fixing $\alpha_{p}$ and
$T_{p}$, can guarantee indefinite saturation of coherence in
principle.  Moreover, since $\chi_{[p]^{m}}\le 2\chi_{[p]^{\infty}}$ for all
$m$, the emergence of coherence saturation in the infinite-time limit 
stroboscopically guarantees high fidelity throughout long storage times.  By
construction, this approach also guarantees that \emph{access latency
is capped} at the duration of the base sequence, with $t_\ell=T_p \ll
T_{s}$; sequence interrupts at intermediate times that are multiples
of $T_p$ are thus permitted in the plateau regime \emph{without
degradation of error suppression}.

Additional insight in the above phenomenon may be gained by recalling
that for free dephasing dynamics ($\alpha_p=0$), the possibility of
non-zero asymptotic coherence is known to occur for supra-Ohmic
($s>1$) bosonic environments \cite{Palma,Hodgson2010}, consistent with
Eq. (\ref{eq:crit}).  The onset of a plateau regime in the controlled 
dynamics may then be given an intuitive interpretation by generalizing the analysis 
carried out in \cite{Hodgson2010} for periodic DD: if the conditions in Eq. (\ref{eq:crit})
are obeyed, the low-frequency (long-time) behavior becomes effectively
supra-ohmic by action of the applied DD sequence and, after a short-time 
transient, the dephasing dynamics ``oscillate in phase'' with the periodically 
repeated blocks.  For sufficiently small $T_p$, the ``differential'' DD error 
accumulated over each cycle in this steady state is very small, leading to the stroboscopic 
plateau.  Interestingly, that phase noise of a local oscillator can saturate at long 
times under suitable spectral conditions has also long been appreciated in the precision
oscillator community~\cite{Rutman}. 

\begin{figure}[t]
\includegraphics[width=\columnwidth]{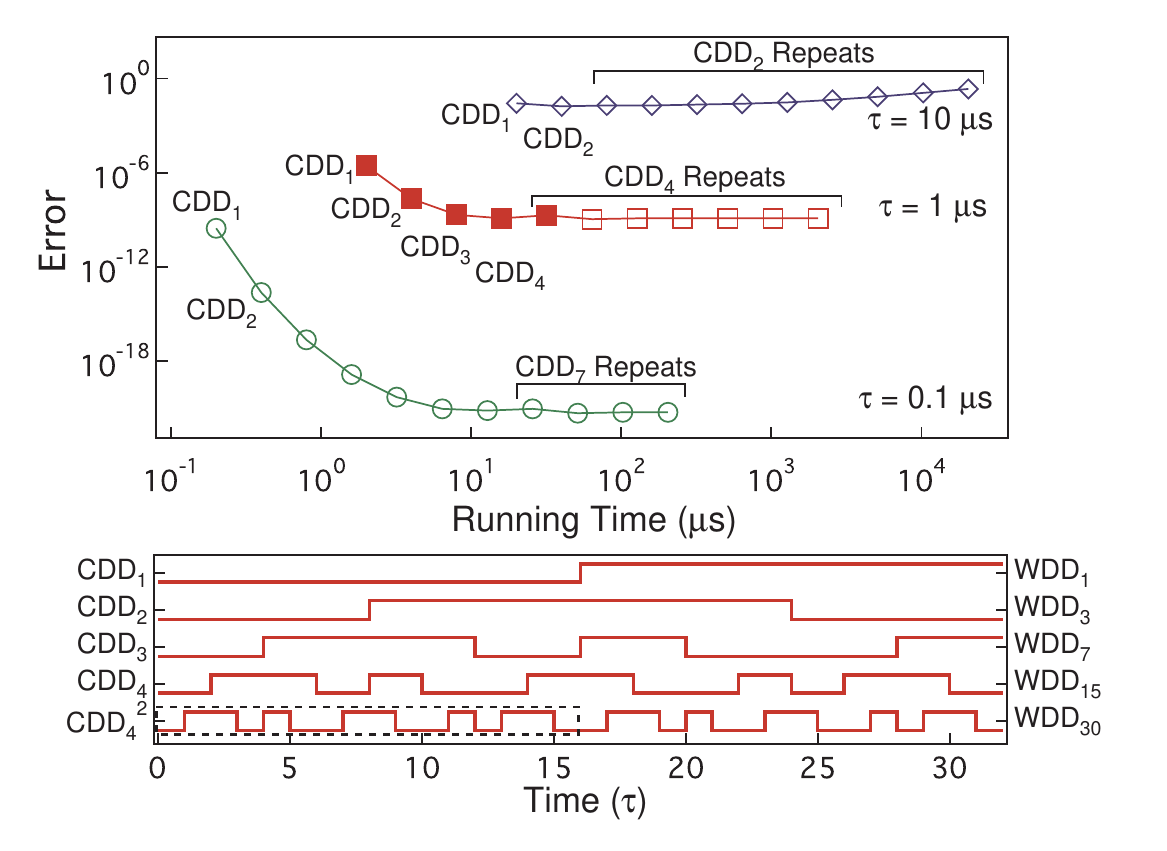}
\vspace*{-6mm}
\caption{{\bf Emergence of coherence plateau and sequence structure.}
Top: Minimal-error DD sequences from numerical search over Walsh DD, for
$\tau = 0.1, 1,10\,\mu$s. 
In each series, the minimal-error sequences systematically access
higher orders of error cancellation (via concatenation) over
increasing running times, until an optimal concatenated sequence is
found which is then repeated in the longer minimal-error sequences.
The gradual increase in error (loss of plateau) for the series with
$\tau=10\,\mu$s is due to the softness of the high-frequency cutoff
and the constraints placed on $T_{p}$ by fixing $\tau$.  For the case
of $\tau=1\;\mu$s, we have calculated the error out to
$m\approx10^{8}$ repeats ($T_s \approx 10^3\,$s, data not shown)
without an observable effect from the soft cutoff.  Bottom: Control
propagators corresponding to the solid markers in the middle data
series ($\tau = 1 \,\mu$s), showing the emergence of a periodic
structure for sufficiently long storage time.  Labels indicate the
corresponding sequence designations in either the CDD or Walsh basis.
Control propagators scaled to same length for ease of comparison.
Dashed box highlights base sequence CDD$_4$ that is repeated for long
times.
\label{fig:wbadd}}
\end{figure}

In light of the above considerations, the occurrence of a
coherence plateau may be observed even for sub-Ohmic noise spectra
($s<1$), as typically encountered, for instance, in both
spin qubits ($s=-2$, as in Fig. \ref{fig:latency}) and
trapped ions ($s=-1$, \cite{Monroe_RMP}).  Numerical calculations of
the DD error using such realistic noise spectra
demonstrate both the plateau phenomenon and the \emph{natural}
emergence of periodically repeated sequences as an efficient solution
for long-time storage, also confirming the intuitive picture given
above.  In these calculations, we employ a direct bandwidth-adapted DD 
search up to time $T_s$, by enforcing additional {\em sequencing constraints}.
Specifically, we turn to Walsh DD, wherein pulse patterns are
given by the Walsh functions, to provide solutions that are efficient
in the complexity of sequencing~\cite{HKBV2011}.  Walsh DD comprises familiar DD protocols, such
as spin echo, Carr-Purcell, and CDD, along with more general protocols,
including repetitions of shorter sequences.  

Starting with a free evolution of duration $\tau$, all possible Walsh DD sequences can be
recursively built out of simpler ones within Walsh DD, doubling in length with
each step.  Further, since all interpulse intervals in Walsh DD protocols re constrained to be 
integer multiples of $\tau$, there are $\frac{1}{2} (T_s/\tau)$ Walsh DD sequences that stop at
time $T_s$, a very small subset of all $2^{T_s/\tau}$ possible digital
sequences, enabling an otherwise intractable bandwidth-adapted DD 
numerical minimization of the spectral overlap integral $\chi_p$.

Representative results are shown in Fig. \ref{fig:wbadd}, where for
each $T_s$ all Walsh DD sequences with given $\tau$ are evaluated and those
with the lowest error are selected.  The choice of $\tau$ sets the
minimum achievable error and also determines whether a plateau is
achievable, as, for a given $T_{s}$, it influences the available values of $T_{p}$ and
$\alpha_{p}$.  As $T_s$ grows, the best performing
sequences (shown) are found to consist of a few concatenation steps
(increasing $\alpha_{p}$ of the base sequence to obey
Eq.~(\ref{eq:crit})), followed by successive repetitions of that fixed
cycle. 
Once the plateau is reached, increasing the number of repetitions does
not affect the calculated error, indicating that stroboscopic sequence 
interrupts would be permitted without performance degradation.
Beside providing a direct means of finding high-fidelity long-time DD
schemes, these numerical results support our key analytic insights as
to use of periodic sequence design.

\vspace*{2mm}

\noindent
{\bf Realistic effects.}  For clarity, we have thus far relied on a
variety of simplifications, including an assumption of pure phase
decoherence and perfect $\pi$ rotations.  However, as we next show,
our results hold in much less idealized scenarios of interest to
experimentalists.  We begin by considering realistic control
limitations.
Of greatest importance is the inclusion of errors due to finite pulse
duration, as they will grow with $T_{s}$ if not appropriately
compensated.  Even starting from the dephasing-dominated scenario we
consider, applying real DD pulses with duration $\tau_{\pi}>0$
introduces both dephasing and depolarization errors, the latter along,
say, the $y$-axis if control along $x$ is used for pulsing.  As a
result, the conditions given in Eq. (\ref{eq:crit}) can no longer
guarantee a coherence plateau in general: simply incorporating
``primitive'' uncorrected $\pi$-pulses into a high-order DD sequence
may contribute a net depolarizing error substantial enough to make a
plateau regime inaccessible.  This intuition may be formalized, and
new conditions for the emergence of a coherence plateau determined, by
exploiting a \emph{generalized multi-axis FF
formalism}~\cite{GreenFF,ToddNJP}, in which both environmental and
finite-width errors may be accounted for, to the leading order, by
adding in quadrature the $z$ and $y$ components of the ``control
vector'' that are generated in the non-ideal setting (see Methods).

The end result of this procedure may be summarized in a transparent
way: to the leading order, the total FF can be written as $F(\omega)
\equiv F_p(\omega) + F_{\text{pul}}(\omega)\approx |A_{\text{bb}}|^{2}
\omega^{2(\alpha_{p}+1)}+|A_{\text{pul}}|^{2}\omega^{2(\alpha_{\text{pul}}+1)}$,
where $F_p(\omega)$ is the FF for the bang-bang DD sequence previously
defined and $F_{\text{pul}}(\omega)$ depends on the details of the
pulse implementation.  Corrections in the pre-factors
$A_{\text{bb}}, A_{\text{pul}}$ arise from higher-order
contributions.  The parameter $\alpha_{\text{pul}}$ captures the error suppression
properties of the pulses themselves, similar to the sequence order of error suppression
$\alpha_p$.  A primitive pulse results in $\alpha_{\text{pul}}=1$ due to the dominant uncorrected
$y$-depolarization.  An expression for the asymptotic DD error may
then be obtained starting from Eq. (\ref{eq:contrep}) and separating
$\chi_{[p]^\infty}\equiv \chi_{[p]^\infty}^{\text{bb}} +
\chi_{[p]^\infty}^{\text{pul}}$.  An additional constraint thus arises
by requiring that \emph{both} the original contribution
$\chi_{[p]^\infty}^{\text{bb}}$ of Eq. (\ref{eq:plateau}) and
$\chi_{[p]^\infty}^{\text{pul}}$ be finite.  Thus, in order to
maintain a coherence plateau in the long-time limit we now require
\begin{equation}
s+2\alpha_{p}>1,\;\; s+2\alpha_{\text{pul}} >1,\;\;T_{p} \omega_{c}< 2 \pi.
\label{eq:critgen}
\end{equation}

\begin{figure}[tb]
\includegraphics[width=0.90\columnwidth]{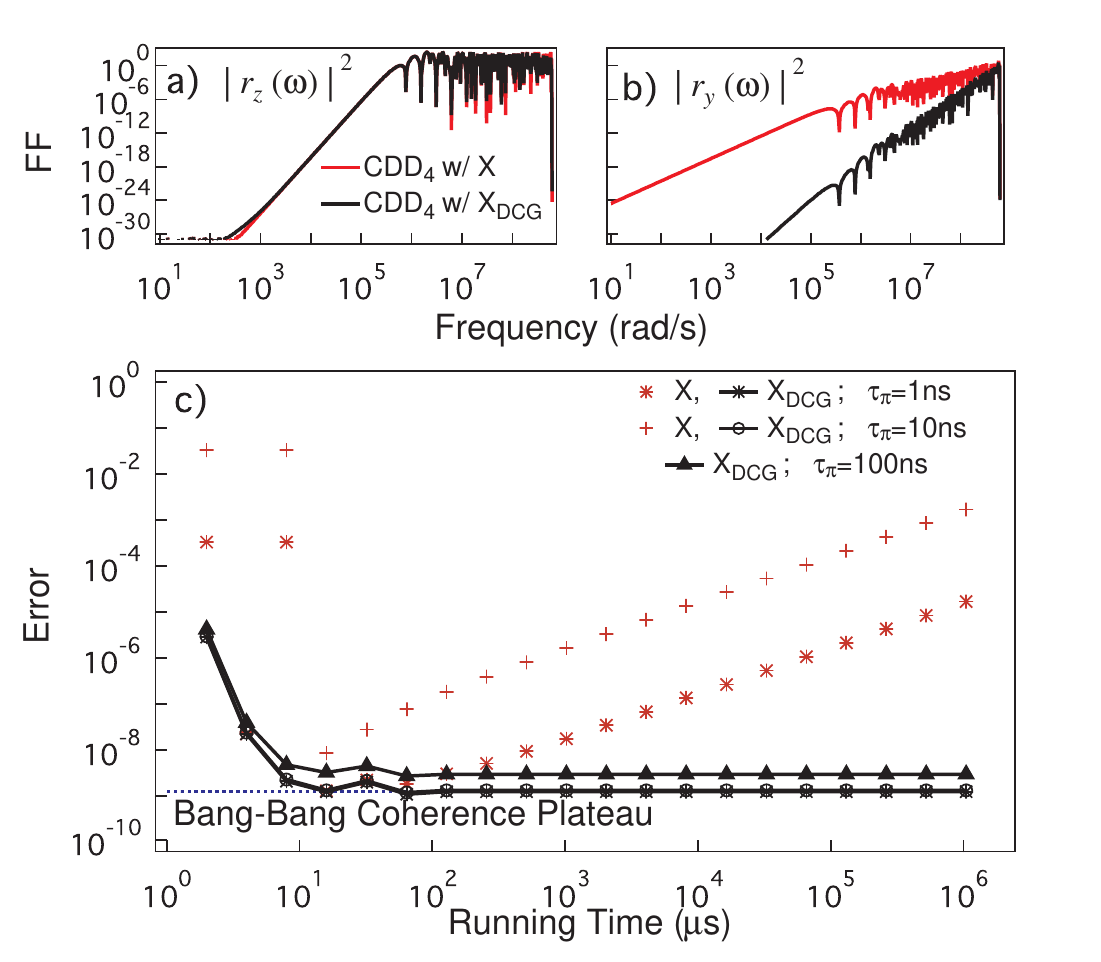}
\vspace*{-2mm}
\caption{{\bf Realistic filter functions and effect of finite-width
errors and soft cutoff.}  a) $z$ (dephasing) and b) $y$
(depolarization) quadrature components of the total FF for CDD$_4$,
$F(\omega)= F_p(\omega) + F_{\text{pul}}(\omega) \equiv
|r_y(\omega)|^2 + |r_z(\omega)|^2$, incorporating non-zero duration
uncorrected $\pi_x$ pulses (red), and first-order DCGs
\cite{khodjasteh_dynamically_2009,ToddNJP}, $\tau_{\pi}=1\,$ns (see also
Methods). In the ideal case, the depolarizing contribution
$|r_y(\omega)|^2\equiv 0$, and $F(\omega) \equiv F_p(\omega)$.  The
improvement of $\alpha_{\text{pul}}$ for CDD$_4$ with DCGs is
demonstrated by the increased slope of $|r_y(\omega)|^2$ in panel b).
c) DD error for the $\tau=1\,\mu$s data set of Fig.~\ref{fig:wbadd},
using finite-duration pulses.  Sub-Ohmic noise spectrum with $s=-2$
and soft Gaussian cutoff as in Fig. \ref{fig:latency} are assumed.
The low-value of $\alpha_{\text{pul}}$ for primitive pulses leads to
unbounded error growth, terminating the plateau after a small number
of repeats, determined by the ratio of $\tau_{\pi}/\tau$.  Sequences
incorporating DCGs meet the conditions for plateau out to \emph{at
least $1\,$s storage time}, with error increased by a factor of order
unity compared with the bang-bang coherence plateau value, using
$\tau_{\pi}$ up to $100\,$ns.  Outlier data points for CDD$_3$ arise
because of even-odd effects in the FF when including pulse effects.
\label{fig:plateaus}}
\end{figure}

We demonstrate the effects of pulse-width errors 
in Fig.~\ref{fig:plateaus}c.  When using primitive $\pi_x$-pulses
($\alpha_{\text{pul}}=1$), the depolarizing contribution due to
$F_{\text{pul}}(\omega)$ dominates the total value of
$\chi_{[p]^{m}}$.  For the dephasing spectrum we consider, $s=-2$, the
condition for maintenance of a plateau using primitive pulses is not
met, and the total error grows unboundedly with $m$ after a maximum
plateau duration $T_{\text{max}}\equiv m_{\text{max}} T_p$
($m_{\text{max}}$ may be estimated by requiring that
$\chi_{[p]^m}^{\text{pul}} > \chi^{\text{bb}}_{[p]^{m}}$, along lines
similar to those discussed in the Methods section).  The unwanted
depolarizing contribution can, however, by suppressed by appropriate
choice of a higher-order ``corrected'' pulse, such as a DCG
\cite{khodjasteh_dynamically_2009,khodjasteh_arbitrarily_2010}, 
already shown to provide efficient error
suppression in the presence of non-Markovian time-dependent
noise~\cite{GreenFF}.  For a first-order DCG, the dominant error
contribution is canceled, resulting in $\alpha_{\text{pul}}=2$, as
illustrated in Figs.~\ref{fig:plateaus}a)-b); incorporating DCGs into
the base DD sequence thus allows the coherence plateau to be restored.
For small values of $\tau_{\pi}$, the error contribution
$\chi^{\text{pul}}_{[p]^{m}}$ remains small and the plateau error is
very close to that obtained in the bang-bang limit.  Increasing
$\tau_{\pi}$ leads this error contribution to grow, and the plateau
saturates at a new higher value.

``Hardware-adapted'' DCGs additionally provide a means to ensure robustness against control imperfections 
(including rotation-angle and/or off-resonance errors) while incorporating realistic control constraints.  For instance, sequences developed for singlet-triplet spin qubits~\cite{aDCG} can simultaneously achieve insensitivity against nuclear-spin decoherence and charge noise in the exchange control fields, with inclusion of finite timing resolution and pulse rise times.
A quantitative performance analysis may be carried out in principle through
appropriate generalization of the FF formalism introduced above.
Thus, the replacement of low-order primitive pulses with
higher-order corrected pulses provides a straightforward path toward
meeting the conditions for a coherence plateau with realistic DD
sequences.  These insights are also supported by recent DD nuclear
magnetic resonance experiments \cite{Suter1,Suter2}, that have demonstrated
the ability to largely eliminate the effects of pulse imperfections in
long pulse trains.

Another experimentally realistic and important control imperfection 
is limited timing precision.  The result of this form of error is either premature or delayed 
memory access at time $T_s' = mT_p\pm \delta t$, offset relative to the
intended one. Qualitatively, the performance degradation resulting from such
access-timing errors may be expected to be similar to the one suffered
by a high-order DD sequence under pulse-timing errors, analyzed
in~\cite{MikeFF}.  A rough sensitivity estimate may be obtained 
by adding an uncompensated ``free-evolution'' period of duration
$\delta t$ following the $m$th repeat of the sequence, with the
resulting FF being determined accordingly. In this case the effective order of suppression
transitions $\alpha_p\to0$, appropriate for free evolution, at a crossover frequency 
determined by the magnitude of the timing jitter.
In order to guarantee the desired (plateau) fidelity level, it is
necessary that the total FF -- including timing errors -- still meets
the requirements set in Eq.~(\ref{eq:critgen}).  In general, this is achievable for 
supra-Ohmic spectra with $s>1$.  When these conditions are not met, the resulting error 
can be much larger than the plateau value if the jitter is appreciable.  
Access timing therefore places a constraint on a system designer to ensure
that quantum memories are clocked with low-jitter, high-resolution
systems.  Considering the situation analyzed in Fig.~\ref{fig:wbadd}
with $\tau=1\,\mu$s and $\chi_{[p]^{\infty}}\sim 1.3\times 10^{-9}$,
we estimate that access jitter of order $1.5\,$ps may be tolerated
before the total measured error exceeds the bound of
$2\chi_{[p]^{\infty}}$. 
Since current digital delay generators allow for sub-ps timing
resolution and ps jitter, the requisite timing accuracy is nevertheless 
within reach with existing technologies.

We next address different aspects of the assumed noise model.
Consider first the assumption of a hard spectral cutoff in bounding
the long-storage-time error.  If such an assumption is not obeyed
(hence residual noise persists beyond $\omega_c$), it is impossible to
fully avoid the singular behavior introduced by the periodic
modulation as $m\rightarrow \infty$.  Contributions from the
resonating region $\omega \approx \omega_{\text{res}}$ are amplified
with $m$, and, similar to pulse-errors, cause $\chi_{[p]^m}$ to
increase unboundedly with time and coherence to ultimately decay to
zero.  Nonetheless, a very large number of repetitions,
$m_{\text{max}}$, may still be applied before such contributions
become important (note that this is the case in the previous figures,
where we assume a {\em soft Gaussian cutoff}).  We lower-bound
$m_{\text{max}}$ by considering a scenario in which a plateau is
preserved with a hard cutoff and estimating when contributions to
error for frequencies $\omega>\omega_{c}$ become comparable to the
plateau error.  For simplicity, 
we assume that noise for $\omega>\omega_{c}$ falls in the
passband of the FF and that at $\omega=\omega_{c}$ the noise power law
changes from $\omega^s \to \omega^{-r}$, with $r>0$.  Treating such a
case with $s=-2$ and using again repeated CDD$_4$ with $\tau=1\;\mu$s
as in Fig.~\ref{fig:wbadd}, we find that as long as $r$ is
sufficiently large, the plateau error $\chi_{[p]^{\infty}}\sim10^{-9}$
can persist for $m_{\text{max}}\gtrsim 10^4$-$10^6$ repetitions (that
is, up to a storage time of over $10\,$s), 
before the accumulated error due to high-frequency contributions
exceeds the plateau coherence (see Methods).  
This makes it possible to engineer a coherence plateau over an
intermediate range of $T_{s}$ which can still be exceptionally long from a
\emph{practical} standpoint, depending on the specific rolloff
behavior of $S(\omega)$ at frequencies beyond $\omega_c$.

Lastly, we turn to consideration of more general open-system models.
For instance, consider a system-bath interaction which includes both a
dominant dephasing component and an ``off-axis'' perturbation,
resulting in energy relaxation with a characteristic timescale $T_1$.
Then the initial dephasing dynamics, \emph{including the onset of a
coherence plateau}, will not be appreciably modified so long as these
two noise sources are uncorrelated and there is a sufficient
separation of time scales.  If $T_1 \gg T_2$, and the maximum error
{\em per cycle} is kept sufficiently small, the plateau will persist
until uncorrected $T_{1}$ errors dominate $\chi_{[p]^{m}}$.  We
reiterate that in many experimentally relevant settings - notably,
both trapped-ion and spin qubits - 
$T_{1}$ effects may indeed be neglected up to very long storage times.
Ultimately, stochastic error sources due, for instance, to spontaneous
emission processes and/or Markovian noise (including white control
noise) may form a limiting mechanism.  In such circumstances, the
unfavorable exponential scaling of Markovian errors with storage time poses a 
problem for high-fidelity storage through DD alone.  
Given a simple exponential decay with time-constant $T_{\textrm{M}}$ and assuming that 
Eq.~(\ref{eq:critgen}) is met, we may estimate a maximum allowed
plateau duration as $T_{\text{max}}\approx T_{\textrm{M}}\chi_{[p]^{\infty}}$.
Thus, even with $T_{\textrm{M}}=100\,$s, a plateau at 
$\chi_{[p]^{\infty}}=10^{-5}$ would terminate after
$T_{\text{max}}=1\,$ms.
Our results thus confirm that guaranteeing high-fidelity quantum
memory through DD alone requires Markovian noise sources
to be minimized, or else motivates the combination of our approach with quantum error
correction protocols.

\vspace*{2mm}

\noindent 
{\bf Discussion} 
\par\noindent 
The potential performance provided by our approach is quite
remarkable.  Besides the illustrative error calculations we have
already presented, we find that many other interesting scenarios arise
where extremely low error rates can be achieved in realistic noise
environments for leading quantum technologies.  For instance,
Ytterbium ion qubits, of direct relevance to applications in quantum
repeaters, allow long-time, low-error coherence plateaus at the
timescale of \emph{hours}, based on bare free-induction-decay ($1/e$)
times of order seconds~\cite{Monroe_RMP}.  Calculations using a common
$1/\omega$ noise power spectrum with CDD$_2$, a Gaussian
high-frequency cutoff near $100\,$Hz, $\tau=1$ms, and DCG operations
with $\tau_{\pi}=10\;\mu$s, give an estimate of the plateau error rate
of $2.5\times 10^{-9}$.  This kind of error rate -- and the
corresponding access latency of just 4 ms -- has the potential to
truly enable viable quantum memories for repeater applications.
Similarly, the calculations shown throughout the manuscript rely on
the well-characterized noise power spectrum associated with nuclear
spin fluctuations in spin qubits.  Appropriate sequence construction
and timing selection \cite{aDCG} permits the analytical criteria set
out in Eq.~(\ref{eq:crit}) to be met, and similar error rates to be
achieved, subject to the limits of Markovian noise processes as
described above.

In summary, we have addressed a fundamental and timely problem in
quantum information processing -- determining a means to effectively produce a
practically useful high-fidelity quantum memory, by using dynamical 
error suppression techniques.  We have identified the key
requirements towards this end, and developed a strategy for sequence
construction based on repetition of high-order DD base sequences.  Our
results allow analytical bounding of the long-time error rates and
identify conditions in which a \emph{maximum error rate can be
stroboscopically guaranteed for long times} with small access latencies, 
even in the presence of limited control.  We have validated these
insights and analytic calculations using an efficient search over Walsh DD
sequences assuming realistic noise spectra.   The results of our numerical 
search bear similarity to an analytically defined strategy established in \cite{Hodgson2010} 
for optimizing long-time storage in a supra-Ohmic excitonic qubit.  

From a practical perspective, our analyses help set technological targets on
parameters such as error-per-pulse, timing resolution, and Markovian
noise strengths required to achieve the full benefits of our approach
to quantum memory.  This work also clearly shows how a system designer
may calculate the impact of such imperfections for a specific
platform, bound performance, and examine technological trade-offs in
attempting to reach a target memory fidelity and storage time.  
As the role of optimization in any particular setting is limited to finding a 
low-error sequence of duration $T_p$ to be repeated up to $T_s$, our 
framework dramatically reduces the complexity of finding high-performance DD protocols.

Future work will characterize the extent to which
similar strategies may be employed to tackle more generic
quantum memory scenarios.  For instance, recent theoretical methods
permit consideration of noise correlations across different spatial
directions~\cite{ToddNJP} in general non-Markovian single-qubit
environments for which $T_2$ and $T_1$ may be comparable.  In such
cases, multi-axis DD sequences such as XY4 \cite{Viola1999Dec} may be
considered from the outset in order to suppress phase and energy
relaxation, as experimentally demonstrated
recently~\cite{Suter_Iterative}.  
Likewise, we remark that our approach naturally applies to
multiple qubits subject to dephasing from {\em independent}
environments. Since expressions similar to the spectral overlap
integral still determine the decay rates of different coherence
elements \cite{Duan1998}, exact DD can be achieved by simply replacing
individual with \emph{collective} $\pi$ pulses, and conditions similar
to Eq. (\ref{eq:plateau}) may then be separately envisioned to ensure
that each coherence element saturates, again resulting in a guaranteed
high storage fidelity.  Addressing the role of correlated dephasing
noise and/or other realistic effects in multi-qubit long-time storage
represents another important extension of this work.

\vspace*{2mm}

\noindent
{\bf Methods} \\
\noindent
{\bf Inclusion of pulse errors.}  Consider a base sequence $p$ of
total duration $T_p$, including both free evolution periods and
control pulses with non-zero duration $\tau_\pi$, where the center of
the $j$th pulse occurs at time $t_j \equiv \delta_j T_p$, with
$\delta_j \in [0, 1]$.  FFs that incorporate, to leading order in
$T_p$, errors due to both dephasing dynamics and non-ideal pulses are
derived following \cite{ToddNJP}.  The total FF, $F(\omega) =
F_p(\omega) + F_{\text{pul}}(\omega)$, may be expressed as
\begin{equation}
F (\omega)\equiv |r_{y}(\omega)|^2+|r_{z}(\omega)|^2 ,
\label{eq:methods1}
\end{equation}
where $r_{z(y)}$ are, respectively, the total $z(y)$ components of the
control vector for pure dephasing in the relevant quadrature,
determined by the toggling-frame Hamiltonian associated with the
control sequence. In the ideal bang-bang limit, $r_y(\omega)\equiv 0$
and $r_z(\omega)= A_{\text{bb}} \omega^{\alpha_p +1}$, where {\em
e.g.}  $\alpha_p =4$, $A_{\text{bb}}= -i T_p^5/2^{14}$ for CDD$_4$.
In general, the total contributions to the FF are
\begin{eqnarray}
r_{z}(\omega) &\hspace*{-1mm} = \hspace*{-1mm} & 
1-e^{i\omega T_p}+\left[2\cos{(\omega\tau_{\pi}/2)}-
e^{-i\omega\tau_{\pi}/2}r_{z}^{\text{pul}}(\omega)\right] u_p ,
\nonumber \\
r_{y}(\omega) & \hspace*{-1mm} = \hspace*{-1mm} & 
-e^{-i\omega\tau_{p}/2}r_{y}^{\text{pul}}(\omega)u_p ,
\end{eqnarray}
\noindent 
where $ u_p \equiv
\sum_{\ell=1}^{n}(-1)^{\ell}e^{i\omega\delta_{\ell} T_p}$ and we
incorporate pulse contributions through $r_{z(y)}^{\text{pul}}$.

For primitive pulses with a rectangular profile, and
$\Omega\equiv\pi/\tau_{\pi}$, direct calculation yields
\cite{GreenFF}:
\begin{eqnarray}
&r^{\text{pul}}_{z} (\omega)=
\frac{\omega^{2}}{(\omega^{2}-\Omega^{2})}
\left(e^{i\omega\tau_{\pi}}+1\right),\nonumber \\
&r^{\text{pul}}_{y} (\omega)=\frac{i\omega\Omega}{(\omega^{2}-\Omega^{2})}
\left(e^{i\omega\tau_{\pi}}+1\right).
\end{eqnarray}
\noindent 
For the 3-segment first-order DCG we employ, one finds instead
\cite{GreenFF, ToddNJP}:
\begin{eqnarray}
r^{\text{pul}}_{z} (\omega)=\omega^{2}\left[\frac{c_{1}(\omega)}
{(\omega^{2}-\Omega^{2})}-\frac{c_{2}(\omega)}
{(\omega^{2}-(\Omega/2)^{2})}\right] , \nonumber \\
r^{\text{pul}}_{y} (\omega)=i\omega\Omega\left[\frac{c_{1}(\omega)}
{(\omega^{2}-\Omega^{2})}-\frac{c_{2}(\omega)}
{2(\omega^{2}-\left(\Omega/2\right)^{2})} \right],
\end{eqnarray}
\noindent 
where $c_{1}(\omega)\equiv
e^{4i\omega\tau_{\pi}}+e^{3i\omega\tau_{\pi}}+e^{i\omega\tau_{\pi}}+1$
and $c_{2}(\omega)\equiv e^{3i\omega\tau_{\pi}}+
e^{i\omega\tau_{\pi}}$.  Starting from these expressions and suitably
Taylor-expanding around $\omega = 0$, one may then show that the
dominant pulse contributions arise from $r_y(\omega)$ in the
uncorrected case, with $\alpha_{\text{pul}}=1$ and $A_{\text{pul}} =
-T_p \tau_\pi/\pi$, whereas they arise from $r_z(\omega)$ in the DCG
case, with $\alpha_{\text{pul}}=2$ and $A_{\text{pul}} = -2 i T_p
\tau_\pi^2/(1+1/\pi^2)$.

Assuming a noise power spectrum with a hard cutoff, $S(\omega) = g
(\omega/\omega_c)^s \Theta (\omega- \omega_c)$, the following
expression for the (leading-order) total asymptotic DD error,
$\chi_{[p]^\infty}\equiv \chi_{[p]^\infty}^{\text{bb}} +
\chi_{[p]^\infty}^{\text{pul}}$, is obtained:
\begin{equation}
\chi_{[p]^\infty} =
\frac{g|A_{\text{bb}}|^{2}\omega_{c}^{2\alpha_p -1}}{\pi
  T_{p}^{2}(s+2\alpha_{p}-1)}+
\frac{g|A_{\text{pul}}|^{2}\omega_{c}^{2\alpha_{\text{pul}}-1}}{\pi
  T_{p}^{2}(s+2\alpha_{\text{pul}}-1)},
\label{eq:methods2}
\end{equation}
leading to the plateau conditions quoted in Eq. (\ref{eq:critgen}).

\vspace*{1mm}

\noindent 
{\bf Effect of a soft spectral cutoff.}  Consider, again, a high-order
DD sequence which is implemented with realistic pulses and is repeated
$m$ times.  Then the leading contribution to the DD is given by
\begin{equation}
\chi_{[p]^{m}}=\int_{0}^{\infty}\frac{S(\omega)}{2\pi\omega^{2}}
\frac{\sin^{2}(m\omega T_{p}/2)}{\sin^{2}(\omega T_{p}/2)}F(\omega) 
d\omega ,  
\label{eq:error}
\end{equation}
where the FF $F(\omega)$ is computed as described above and $S(\omega)
= g (\omega/\omega_c)^s f(\omega, \omega_c)$.  While this integral
converges nicely if we assume a sharp high-frequency cutoff, this is
rarely encountered in reality.  For a soft spectral cutoff, we can
break the error integral up into two (low- vs. high-frequency)
contributions, say, $ \chi_{[p]^{m}} \equiv
\chi_{[p]^{m}}^{\text{low}} + \chi_{[p]^{m}}^{\text{high}}$.  We wish
to estimate how many repeats of the base sequence are permitted
\emph{under conditions otherwise leading to a plateau}, before
corrections due to the high-frequency tail dominate the error behavior
and destroy the plateau.  Assume that the conditions given in
Eq. (\ref{eq:critgen}) are obeyed, and let the maximum number of
allowed repetitions be denoted by $m_{\text{max}}$.  Then
$m_{\text{max}}$ may be determined by requiring that
$\chi_{[p]^{m_\text{max}}}^{\text{low}} =
\chi_{[p]^{m_\text{max}}}^{\text{high}}$.

Since, for every $m$, we have $\chi_{[p]^{m}}^{\text{low}} \leq 2
\chi_{[p]^{\infty}}^{\text{low}}$, a lower bound for $m_{\text{max}}$
may be obtained by estimating $m_\ast$ such that
$\chi_{[p]^{m_\ast}}^{\text{high}} = 2
\chi_{[p]^{\infty}}^{\text{low}}$.  We may therefore simply identify
$\chi_{[p]^{\infty}}^{\text{low}}$ with the hard-cutoff asymptotic
value given in Eq. (\ref{eq:methods2}).  In order to obtain an
explicit expression for the high-frequency contribution, we assume
that the noise power above $\omega_{c}$ also takes a power-law form,
$S(\omega)= g (\omega/\omega_c)^r$, formally corresponding to a
rolloff $f=(\omega/\omega_c)^{r-s}$, with power $r>0$.  
(Note that other possible choices of $f$, such as exponential or
Gaussian rolloffs, may be treated along similar lines, at the expense
of more complicated integrals). Thus, we may write
\begin{eqnarray}
\chi_{[p]^{m_\ast}}^{\text{high}} \leq \hspace*{-1mm}
\int_{\omega_{c}}^{\infty}\frac{g
(\omega/\omega_c)^{-r}}{2\pi\omega^{2}} \frac{\sin^{2}(m\omega
T_{p}/2)}{\sin^{2}(\omega T_{p}/2)} F_{[p]}^{\text{max}}
d\omega ,
\end{eqnarray}
\noindent 
where we have set the FF to the maximum value
$F_{[p]}^{\text{max}}\equiv F_{[p]}^{\text{max}}(n, \alpha)$ of the
peaks in the passband.  This value increases with pulse number and
sequence order and must be calculated explicitly for a particular base
sequence.

For sufficiently large $m$, the oscillatory factor in the integrand
may be approximated in terms of a Dirac comb,
\begin{equation}
\frac{\sin^{2}(m\omega T_{p}/2)}{\sin^{2}(\omega T_{p}/2)} \approx
\frac{2\pi m}{T_p} \sum_{n=-\infty}^{\infty}\delta\left(\omega-\frac{2\pi
n}{T_{p}}\right). 
\end{equation}
This allows us to write
\begin{eqnarray}
\chi_{[p]^{m_\ast}}^{\text{high}} & \hspace*{-1mm}\lesssim \hspace*{-1mm}& 
F_{[p]}^{\text{max}}(n,\alpha_p) \frac{g \omega_c^{r}}{T_p}
\sum_{n=1}^{\infty}\left(\frac{2\pi n}{T_{p}}\right)^{-(r+2)} 
\nonumber \\ 
& \hspace*{-1mm}=  \hspace*{-1mm}&
\frac{m g T_p F_{[p]}^{\text{max}}(n,\alpha_p) }{4 \pi^2} \left
(\frac{\omega_c T_p}{2\pi} \right)^r \hspace*{-1.5mm} \zeta (r+2) ,
\end{eqnarray}
\noindent 
where we have exploited the fact that $0< \omega_c < 2\pi/T_p$ and
$\zeta(s)$ denotes the Riemann zeta function.

The error due to the soft rolloff at high frequencies thus increases
linearly with $m$ (hence $T_s=m T_p$), as intuition suggests.  Since
the zeta function is decreasing with $r$ and attains its maximum value
at $r=0$, corresponding to an infinite white noise floor, we obtain
the following upper bound (recall that $\zeta(2)=\pi^{2}/6$):
\begin{equation}
\chi_{[p]^{m_\ast}}^{\text{high}} \lesssim \frac{1}{24} m g T_p
F_{[p]}^{\text{max}}(n,\alpha_p) \left(\frac{\omega_c T_{p}}{2\pi}
\right)^r .
\label{eq:methods3}
\end{equation}
By equating $\chi_{[p]^{m_\ast}}^{\text{high}} = 2
\chi_{[p]^{\infty}}^{\text{low}}$ and using
Eqs. (\ref{eq:methods2})-(\ref{eq:methods3}), we finally arrive at the
desired lower-bound:
\begin{equation}
m_{\text{max}} \gtrsim \frac{48}{ F_{[p]}^{\text{max}}(n,
\alpha_p) } \left( \frac{2\pi}{\omega_c T_p}\right)^{r} 
\left ( \frac{\chi_{[p]^{\infty}}^{\text{low}}}{g T_p}    \right).
\end{equation}

The above estimate can be applied, in particular, to the specific
situation analyzed in the main text: base sequence CDD$_4$ with
$\tau=1\mu$s, DCG implementations with $\tau_\pi \leq 10~$ns, and
$s=-2$.  In this case $T_p\approx 16\mu$s, $\alpha_{p}=4$,
$F_{[p]}^{\text{max}}(n, \alpha_p)=256$, $A_{\text{bb}}=-i
{T_{p}^{5}}/{2^{14}}$, and one can effectively neglect the
contribution to $m_{\text{max}}$ due to pulse errors to within the
accuracy of this lower bound.  Let $x\equiv T_p \omega_c /2\pi$ which,
by the assumed plateau condition, ranges within $[0, 1]$.  Then we may
rewrite
\begin{equation}
m_{\text{max}} \gtrsim \frac{3\pi^6}{5 \times 2^{25}} \ x^{-r+7}, 
\label{eq:bound}
\end{equation}
implying that, for instance, at least $10^5$ repetitions are allowed
at $x=0.001$ if $r \geq 6$, and at least $10^4$ at $x=0.01$ if $r \geq
8$.  At the value $x=0.16$, corresponding to $\omega_c/2\pi$ as used
in the main text, $r \gtrsim 18$ ensures $m_{\text{max}} \gtrsim 10^4$
hence a storage time of about $T_s \approx 0.1\,$s {\em with error as
low as $10^{-9}$}. As demonstrated by the data in
Fig. \ref{fig:plateaus}, $T_s$ is in fact in excess of $1\,$s under
the assumed Gaussian cutoff, which is realistic for this system.  In
general, we have verified by direct numerical evaluation of the error
integral in Eq. (\ref{eq:error}) that, although qualitatively correct,
the lower bound in Eq. (\ref{eq:bound}) can significantly
under-estimate the achievable plateau length ({\em e.g.}, at $x=0.16$,
a storage time $T_s \approx 0.1\,$s is reached already at $r \gtrsim
15$).  Altogether, this analysis thus indicates that high-frequency
tails do not pose a practically significant limitation provided that
the noise falls off sufficiently fast, as anticipated.


%

\vspace*{3mm}

\noindent 
{\bf Acknowledgements}\\
\noindent 
Work supported by the US ARO under contract No. W911NF-11-1-0068, the
US NSF under grant No. PHY-0903727 (to LV), the IARPA QCS program
under contract No. RC051-S4 (to LV), the IARPA MQCO program (to MJB),
and the ARC Centre for Engineered Quantum Systems, CE110001013.  We
thank Seung-Woo Lee for a critical reading of the manuscript.

\vspace*{1mm}

\noindent 
{\bf Author contributions}\\
\noindent 
L.V. formulated the problem.  K.K. and L.V. established analytical error bounds 
and plateau conditions in the ideal case. M.J.B. led the analysis of realistic effects and systematic impacts, 
with FF calculations being carried out by J.S., D.H., and T.J.G. 
All authors jointly validated the results and prepared the manuscript.

\end{document}